# Metallicity of $Ca_2Cu_6P_5$ with Single and Double Copper-Pnictide Layers

Li Li[1], David Parker[1], Miaofang Chi,[2] Georgiy M. Tsoi,[3] Yogesh K. Vohra,[3] and Athena S. Sefat[1]

[1] Materials Science & Technology Division, Oak Ridge National Laboratory, Oak Ridge, TN 37831, USA

[2] Center for Nanophase Materials Sciences, Oak Ridge National Laboratory, Oak Ridge, TN, 37831, USA

[3] Department of Physics, University of Alabama at Birmingham, Birmingham, AL 35294, USA

Corresponding author emails: lil2@ornl.gov; sefata@ornl.gov

**Abstract**

We report thermodynamic and transport properties, and also theoretical calculations, for Cu-based compound $Ca_2Cu_6P_5$ and compare with $CaCu_{2-\delta}P_2$. Both materials have layers of edge-sharing copper pnictide tetrahedral $CuP_4$, similar to Fe-As and Fe-Se layers (with $FeAs_4$, $FeSe_4$) in the iron-based superconductors. Despite the presence of this similar transition-metal pnictide layer, we find that both $Ca_2Cu_6P_5$ and $CaCu_{2-\delta}P_2$ have temperature-independent magnetic susceptibility and show metallic behavior with no evidence of either magnetic ordering or superconductivity down to 1.8 K. $CaCu_{2-\delta}P_2$ is slightly off-stoichiometric, with $\delta = 0.14$. Theoretical calculations suggest that unlike Fe $3d$-based magnetic materials with a large density of states (DOS) at the Fermi surface, Cu have comparatively low DOS, with the majority of the $3d$ spectral weight located well below Fermi level. The room-temperature resistivity value of $Ca_2Cu_6P_5$ is only 9 $\mu\Omega$-cm, due to a substantial plasma frequency and an inferred electron-phonon coupling $\lambda$ of 0.073 (significantly smaller than that of metallic Cu). Also, microscopy result shows that Cu-Cu distance along the $c$-axis within the double layers can be very short (2.5 Å), even shorter than metallic elemental copper bond (2.56 Å). The value of $d\rho/dT$ for $CaCu_{2-\delta}P_2$ at 300 K is approximately three times larger than in $Ca_2Cu_6P_5$, which suggests the likelihood of stronger electron-phonon coupling. This study shows that the details of Cu-P layers and bonding are important for their transport characteristics. In addition, it emphasizes the remarkable character of the DOS of '122' iron-based materials, despite much structural similarities.

# 1 Introduction

The discovery of high-$T_C$ superconductivity in LaFeAsO[1] has generated extensive interest in iron-based superconductors and other layered transition metal based materials, whose structures contain MX (M= transition metal, X = pnicogen or chalcogen) layers composed of edge-sharing $MX_4$ tetrahedra[2]. Among the vast number of compounds, the $ThCr_2Si_2$-type $AM_2Pn_2$ pnictides (A = Ca, Sr, Ba, Pn= P, As, Sb) have attracted much attention due to their wide range of doping possibilities either by holes or electrons, as well as the recipe for single crystals with sufficient sizes for various experiments. It is suggested that this type of material is located at the borderline of an interlayer Pn-Pn bonding instability, and the forming and breaking of a Pn-Pn bond will lead to a shape change of the tetrahedron as well as the charge redistribution within the Pn-M-Pn bonded layers[3]. Different choices of A or M element may affect the formation of Pn-Pn bond in these phases, and further influence the crystal and electronic structures, magnetism, and presence of superconductivity in the ternary pnictides. Following this assumption, many efforts have been made to revisited the 122-type pnictides and varieties of interesting behaviors have been found other than the high-temperature superconductivity, e.g., itinerant antiferromagnetism in $BaCr_2As_2$ [4] and $CaCo_2P_2$,[5] antiferromagnetic insulator in $BaMn_2As_2$,[6] Pauli paramagnetic metals in $CaFe_2P_2$ [7] and $CaNi_2P_2$,[7] metal with nearly T-independent diamagnetism in $ACu_2Pn_2$ (A = Sr, Ba; Pn = As, Sb)[8,9], and superconductivity in $BaNi_2P_2$ [10] and $SrNi_2P_2$ ($T_C < 4$ K)[11]. By analogy with the cuprates, it is interesting to consider whether high superconductivity can be observed in material with copper phosphide layers, in a similar way to iron-based superconductors.

Our work here focuses on two calcium copper-pnictide layered compounds of $Ca_2Cu_6P_5$ ("265") and $CaCu_{2-\delta}P_2$ ("122"). Both 265 and 122 have layers of edge-sharing copper pnictide tetrahedral ($CuP_4$), similar to iron arsenides in iron-based superconductors (FeSC). 265 has both single and double P-Cu-P layers, while 122 is only made up of single layers. X-ray refinement result shows there are ~ 7% vacancies on the Cu-sites in $CaCu_{2-\delta}P_2$, which is similarly found in arsenide version of $CaCu_{1.7}As_2$[12]. Through extensive properties measurements, we find no evidence of either magnetic ordering or superconductivity in these two copper pnictides, despite the resemblance of crystal structure building units to the iron-based materials. Both $Ca_2Cu_6P_5$ and $CaCu_{2-\delta}P_2$ show temperature-independent weak magnetic susceptibility ($\chi \approx 10^{-5}$ emu/mol.Oe) and $CaCu_{2-\delta}P_2$ forms with Cu vacancies that lower the Fermi-level DOS and T-linear specific heat coefficient. Our structural details and theoretical calculations explained the metallic behavior in these two compounds and suggest the importance of the details of Cu-P layers on the transport characteristics of these materials.

# 2 Experimental

The elemental reactants, dendritic Ca pieces (99.98%), Cu powders (99.999%), P lumps (99.999%) were all from Alfa Aesar. Prior to its use, copper powder was reduced in a diluted $H_2$ stream (5/96% $H_2$/Ar) for 12 h, whereas other elements were used as received. All operations were handled in a helium-filled glovebox with the moisture and oxygen below 0.1 ppm. CaP binary was prepared by placing mixtures of Ca pieces and P lumps in a silica tube, which were evacuated to a pressure of $10^{-2}$ Torr and placed in a horizon tube furnace. The mixtures were slowly heated up to 400 °C (annealed for 1 day), then heated up to 750 °C (annealed for 3 days), then slowly cooled to room temperature. Cu$_2$P binary was prepared from the mixtures of Cu powders and P lumps by using the same method. Ca$_2$Cu$_6$P$_5$ and CaCu$_{2-\delta}$P$_2$ were grown out of CaP and Cu$_2$P binaries, with the ratio of 2:1 and 1:1, respectively. The reaction mixtures were sealed inside evacuated silica tubes, annealed at 900 °C (for 265) and 950 °C (for 122) for 12 hours, and then slowly cooled to room temperature over 10 hours. The samples were ground and pressed into pellets, sealed under vacuum inside silica tubes, heated to 1000 °C (for 265) and 950 °C (for 122), and annealed for 24 hours. After another regrinding and pelletizing, the pellets were sealed in evacuated silica tubes and annealed at 900 °C (for 265) and 950 °C (for 122) for 48 hours before turning off the furnace. The obtained samples were black and stable in air.

The powder x-ray diffraction (XRD) patterns were collected at room temperature by a PANalytical X'Pert PRO MPD x-ray powder diffractometer with Cu Kα radiation. The Rietveld refinements of the XRD patterns were done with FullProf software suite.[13] The electrical and thermal transport measurements were performed in a Quantum Design (QD) Physical Property Measurement System (PPMS), using a standard four-probe method. Magnetization measurements were performed in a QD Magnetic Property Measurement System (MPMS).

Scanning Transmission Electron Microscopy (STEM) images on 265 were acquired using High Angle Annular Dark Field (HAADF) and Annular Bright Field (ABF) imaging methods in STEM mode with an FEI Titan 60-300S microscope equipped with a Gatan Imaging Filter Quantum 865. The electron probe used in the imaging was ~0.7A at a full-width half-maximum with a convergence semi-angle of 23mrad at 300kV. A collection semi-angle of 55-300mrad for HAADF imaging, and of 11.5- 23mrad for ABF imaging was implemented in this experiment. The TEM specimens were prepared by focused ion beam sectioning with a low ion-beam current followed by nano-milling on a cryo-stage for the final thinning.

High pressure electrical resistance measurements at cryogenic temperatures down to 10 K were performed in a six-tungsten microprobe designer diamond anvil cell (DAC) with 280 μm size culet[14,15]. The tungsten microprobes are encapsulated in a homoepitaxial diamond film and are exposed only on the culet surface to make contact with the sample. The sample was loaded into an 80 μm hole of a 230 μm thick spring-steel gasket that was first pre-indented to a ~80 μm thickness and mounted between the designer diamond anvil and 300 μm culet size matching diamond. Two electrical leads pass constant 2 mA current through the sample and two additional leads measure voltage across the sample. The sample pressure was monitored continuously by the in situ ruby fluorescent technique[16] and care was taken to carefully calibrate the ruby R1 emission to the low temperature of 10 K.[14] In this study no pressure medium was used in order to ensure good electrical contacts between the probes and the sample.

First principles density functional theory calculations were performed using the linearized augmented plane wave code WIEN2K[17], within the Generalized Gradient Approximation of Perdew, Burke and Ernzerhof[18]. Internal coordinates were relaxed until forces were less than 2mRyd/Bohr, and sphere radii of 2.1 Bohr for P and 2.3 Bohr for Ca and Cu were used for Ca$_2$Cu$_6$P$_5$, and 1.96 and 2.5, respectively, for CaCu$_2$P$_2$. An RK$_{max}$ of 9.0 was used, where R is the smallest LAPW sphere radius and K$_{max}$ the largest plane wave vector. For the purposes of computing the plasma frequencies used to understand the resistivity, we used the Boltzmann transport code BoltzTraP[19]. Approximately 1000 k-points in the full Brillouin zone were used for the self-consistent calculations, and approximately 10,000 points were used for the transport calculations.

## 3 Results and discussion

Fig. 1 (a) shows the XRD patterns and Rietveld refinements for $Ca_2Cu_6P_5$ and $CaCu_{2-\delta}P_2$ structures respectively. The reflections can be well indexed with the space group of $I4/mmm$ (No. 139) for both. The schematic images of the crystal structure are shown in the inset. Only trace amount of impurity phases are present in the XRD pattern, which have been taken into account for a refinement that improved the fitting. The lattice parameters are $a = 4.0169(3)$ Å, $c = 24.6634(2)$ Å for $Ca_2Cu_6P_5$, and $a = 4.0223(2)$ Å, $c = 9.6333(1)$ Å for $CaCu_{2-\delta}P_2$. Impurity phase was estimated using the Hill and Howard method[20]. For 265, the only impurity detected was $CuP_2$, refined to less than 3% by mass; for 122, the total content of impurities were less than 6%. For $CaCu_{2-\delta}P_2$, electron density of Cu-site occupancy according to diffraction data is about 0.93, i.e., the vacancy concentration is ≈7% ($= \delta/2$), corresponding to an approximate composition of $CaCu_{1.86}P_2$. Vacancies were similarly reported for $CaCu_{1.7}As_2$[12].

The STEM results of HAADF image of $Ca_2Cu_6P_5$ and its simultaneously acquired ABF are shown Fig.1(b), which are sensitive to heavy and light elements, respectively. By combining these two imaging techniques, the atomic columns of all three elements, Ca, Cu and P, are clearly observed. An alternating single and double Cu-P tetrahedron layers are resolved. Here, the lattice distortions or bonding length modification from the incorporation of different layered configurations are detected. In fact, the interlayer Cu-Cu bond length in the double P-$Cu_2$-P layers is between 0.25 nm and 0.29 nm, well consistent with the x-ray refinement result (≈ 2.65 Å). In fact, the Cu-Cu distances in the double layer seem to alternate from the detailed image analyses; the atomic position are identified using Gaussian fitting (see Fig. 1b). The Cu-Cu distance within $ab$-plane is 2.84 Å ($=a/\sqrt{2}$), slightly larger than reported for $d_{Fe-Fe}$ in $BaFe_2As_2$ (2.80 Å).

Fig. 2 shows the temperature dependence of electrical resistivity $\rho(T)$ for $Ca_2Cu_6P_5$ and $CaCu_{2-\delta}P_2$. Both of them exhibit almost similar metallic behavior over the whole measured temperature range. At room temperature, the electrical resistivity is about 9 μΩ-cm for $Ca_2Cu_6P_5$ and 47 μΩ-cm for $CaCu_{2-\delta}P_2$. As the temperature decreases, the resistivity decreases monotonously down to 1.5 μΩ-cm for $Ca_2Cu_6P_5$ and 19 μΩ-cm for $CaCu_{2-\delta}P_2$ around 20 K, and then keeps almost constant below 20 K. The $\rho(T)$ of $CaCu_{2-\delta}P_2$ does not show any feature down to 2 K, in disagreement with $CaCu_{1.7}As_2$ exhibiting an anomaly (at ~ 45 K) assumed to be associated with ordering of the Cu vacancies[12]. For the single- and double-layer compound of $Ca_2Cu_6P_5$, we performed a pressure study to see if some correlated behavior can be induced within the double layers. The temperature dependence of resistivity below 100 K with applying hydrostatic pressure of 1.4 GPa, 3.7 GPa, 8.9 GPa, and 11.7 GPa, is shown in Fig. 2 inset. Unfortunately, the sample remains metallic without anomalies. Magnetic susceptibility measurements show that $Ca_2Cu_6P_5$ is Pauli paramagnetic, while $CaCu_{2-\delta}P_2$ is diamagnetic with signals of $\chi \sim 10^{-5}$ emu/mol.Oe (not shown), and a Curie tail below due to paramagnetic impurities as observed in diffraction.

Fig. 3(a) illustrates the temperature dependence of the specific heat for $Ca_2Cu_6P_5$ and $CaCu_{2-\delta}P_2$. There is no anomaly in the whole temperature range, confirming no structural transition or magnetic ordering in either system. The plots of $C_p/T$ versus $T^2$ for $Ca_2Cu_6P_5$ and $CaCu_{2-\delta}P_2$ in the temperature range from 1.8 to 10 K are shown in Fig. 3(b). The experimental data points were fitted using the formula $C_p = \gamma T + \beta T^3$, in which the first and the second terms are the electronic and lattice contributions to the heat capacity, respectively. The obtained results from the fitting are $\gamma = 5.64(2)$ mJ/mol-K$^2$, $\beta = 1.34(1)$ mJ/mol-K$^4$ for $Ca_2Cu_6P_5$, and $\gamma = 1.66(3)$ mJ/mol-K$^2$, $\beta = 0.21(2)$ mJ/mol-K$^4$ for $CaCu_{2-\delta}P_2$. The Debye temperatures are estimated to be 265 K for $Ca_2Cu_6P_5$ and 359 K for $CaCu_{2-\delta}P_2$, from the $\Theta_D = (12\pi^4 NR / 5\beta)^{1/3}$, where N is the number of atoms per formula.

The Seebeck coefficients of $Ca_2Cu_6P_5$ and $CaCu_{2-\delta}P_2$ are shown in Fig. 4(a). The thermopower values increase monotonically within the whole temperature range, achieving about 9.67 µV K$^{-1}$ for $Ca_2Cu_6P_5$ and 4.84 µV K$^{-1}$ for $CaCu_{2-\delta}P_2$, which are both relatively low for metallic compounds. Their thermopowers are positive, indicating that holes are the major charge carriers in the system. Fig. 4(b) shows the temperature dependence of the thermal conductivity κ for $Ca_2Cu_6P_5$ and $CaCu_{2-\delta}P_2$. The thermal conductivity of $Ca_2Cu_6P_5$ rapidly increases to ~100 WK$^{-1}$m$^{-1}$ with a linear behavior from 2 K to 30 K, and after that it is temperature independent up to 300 K. For comparison, the κ value of $Ca_2Cu_6P_5$ at room temperature is close to 109 WK$^{-1}$m$^{-1}$ of brass (~70%Cu-30%Zn)[21], but is only one fourth of 401 WK$^{-1}$m$^{-1}$ for the value of pure copper[21]. For $CaCu_{2-\delta}P_2$, the thermal conductivity value (19 WK$^{-1}$m$^{-1}$) at 300 K is much lower than $Ca_2Cu_6P_5$. The thermal conductivity also shows a rapid increase from 2 K to 30 K ($\kappa_{30K} \approx 11$ WK$^{-1}$m$^{-1}$), and after that it slowly increases with warming up to 300 K.

As mentioned above, the room temperature resistivity value of $Ca_2Cu_6P_5$ is 9 µΩ-cm, which is exceptionally low and may be a result for having small Cu-Cu intraplanar distance of 2.5 Å. By way of comparison, the review of resistivity of cubic metallic elements by Allen[22] shows values at 295 K ranging from 1.61 µΩ-cm for Ag to 39 µΩ-cm for Ba, and fully *half* these elements exceed the 9 µΩ-cm value. In order to understand the high level of metallicity for $Ca_2Cu_6P_5$, we performed first principles DOS calculations. Fig. 5 shows a DOS comparison of 265 to 122. For 265, the greatest DOS is located between 3 and 5 eV below the Fermi level. This is unlike 3d-based magnetic materials in which there is typically a large DOS at the Fermi surface and in fact bears some resemblance to metallic Cu itself. The Fermi-level DOS is just 2.38 states/eV-f.u., and much of the spectral weight within the spheres is Cu, providing the beginnings of an explanation for the low resistivity. For $Ca_2Cu_6P_5$, when we convert the Fermi-level DOS to a T-linear electronic specific heat coefficient γ, we find a γ of 5.62 mJ mol$^{-1}$ K$^{-2}$. This compares very well to the experimental value of 5.64 mJ mol$^{-1}$ K$^{-2}$ and this agreement is in fact suggestive of very weak electron-phonon coupling.

The standard theory of electron-phonon coupling[22] tells us that electron-phonon coupling enhances "bare" band structure values of γ by a factor $(1 + \lambda)$, where λ is the dimensionless electron-phonon coupling constant. As pointed out by Allen, typically $(1 + \lambda)$ is uncertain by ~10%, so that the theory and experiment together here should not be taken to imply a λ of 5.64/5.62 – 1 = 0.0036, but are suggestive of a λ of order 0.1 or less, which is comparable to that of Cu itself. To put this on still more solid computational foundation, we consider that the resistivity ρ in this regime is given by the following expression: $\rho = 8\pi^2 k_b T\lambda/\hbar\omega_p^2$, where $\omega_p$ is the plasma frequency and $k_B$ is the Boltzmann's constant. Now, $\omega_p^2$ is directly calculable from the electronic structure results (it is effectively identified with $\sigma / \tau$ output from the BoltzTraP results; here τ is an average scattering time). There is one slight complication in that in $Ca_2Cu_6P_5$ the plasma frequency is anisotropic due to the tetragonal symmetry; we address this by modification $\omega^2_{p,total}=(2/3) \omega^2_{p,planar} + (1/3) \omega^2_{p,c}$ (this is equivalent to assuming an isotropic scattering time). This yields an $\omega_{p,total}$ of 3.14 eV. Again, this is a comparatively large value; it is comparable to the smallest plasma frequency in the cubic metals compiled by Allen[22], thus forming another piece of the explanation of the low resistivity of $Ca_2Cu_6P_5$. We note that the Fermi surface of this material (depicted in Fig. 5a) shows a large area, consistent with the large plasma frequencies. The resistivity value can be fitted using the first principles calculated plasma frequencies, finding an electron-phonon coupling $\lambda$ in $Ca_2Cu_6P_5$ of 0.073. This is significantly smaller than even the value of 0.13 estimated for metallic Cu, and it is this very small value that is mainly responsible for the surprisingly small resistivity.

Note that we consider the resistivity as being due to electron-phonon coupling, as is usually the case in metals. It is conceivable, though rather unlikely, that there is a contribution from spin-fluctuation scattering arising from magnetic fluctuations in the Cu atoms. No evidence of a magnetic phase transition was seen, either in the experiment (ambient and even pressure work) or in our theoretical calculations (we checked both ferromagnetic and a nearest-neighbor antiferromagnetic state). It is of interest to consider the theoretical results on this layered structure in light of the literature work on the superconducting iron-based materials. Unlike in these materials, here we find no evidence of either magnetic ordering, despite the presence of a Cu-P layer that bears a certain resemblance to the Fe-As or Fe-Se layers in the iron-based materials. These iron-based materials are near to itinerant antiferromagnetism[23], due largely to the presence of the Fe *3d* states at the Fermi level, along with the favorable Fermi surface topology. In the 265 compound, however, as shown in Fig. 5a, the highest density of Cu states are generally far from the Fermi level, as in Cu itself; the Cu Fermi level density of states here is approximately 0.2/Cu-eV, while the calculated Fe Fermi-level DOS in LaOFeAs is approximately 2/Fe-eV, or ten times higher. Hence there is not sufficient spectral weight near $E_F$ in $Ca_2Cu_6P_5$ to drive a magnetic instability. The comparatively low DOS, but substantial plasma frequency, in this material indicates large Fermi velocities. It is interesting that the Fermi surface structure of $Ca_2Cu_6P_5$ contains nearly one dimensional sheets, running vertically and suggestive of nesting, although we find no indication of a magnetic or structural instability in this compound. It is likely the lack of spectral weight at $E_F$, as manifested by the low Fermi-level DOS, precludes such a possibility.

Depicted in Fig. 5b is the electronic structure of $CaCu_{2-\delta}P_2$ that bears some similarity to that of $Ca_2Cu_6P_5$; in both cases there is a comparatively low Fermi-level DOS and the majority of the Cu spectral weight is located well below the Fermi level. The Fermi level DOS in the 122 compound is calculated to be 1.39 states/eV.formula-unit, which translates to a T-linear specific heat coefficient $\gamma = 3.26$ mJ/mol-$K^2$. This is nearly double the experimental value of 1.66 mJ/mol-$K^2$, but we note that the electronic DOS in Fig. 5b varies quite rapidly around $E_F$. It should be noted that our XRD refinements place the occupancy of the Cu site as 0.93, and these vacancies would tend to move the Fermi level leftward on the graph, to regions of lower DOS and hence T-linear specific heat coefficient. To check this we constructed a 2x1x1 supercell of the 122 structure, with one Cu removed, for a total of 4 units of $CaCu_{1.75}P_2$, and calculated its DOS (black dotted line in Fig.5b DOS plot, inset). It is evident that Cu vacancies hole dope the system and lower the Fermi-level DOS; one calculates a $\gamma$ of 1.76 mJ/mol-$K^2$, in much better agreement with the experimental value. The Fermi surface structure (Fig. 5b) is more three-dimensional than 265, which consists mainly of vertical and horizontal sheets. Here there are tube-like and globular structures, at the zone center and edges respectively. Despite this, the directionally averaged plasma frequency is quite similar to that of the 265 material. The Cu vacancies found in the 122 structure may play a role in higher resistivity, with the residual resistivity being an order of magnitude larger. Nevertheless, the slope d$\rho$/dT of the resistivity at 300 K is approximately three times larger than in 265 compound, which given the similarity in plasma frequencies suggests the likelihood of stronger electron-phonon coupling. This illustrates the importance of the number of Cu-P layers on the detailed transport characteristics of these materials.

# 4 Conclusions

In summary, we have synthesized two calcium copper-pnictide layered compounds, $Ca_2Cu_6P_5$ and $CaCu_{2-\delta}P_2$ and systematically studied their thermodynamic and transport properties. Through properties measurement, we find metallicity for both $Ca_2Cu_6P_5$ and $CaCu_{2-\delta}P_2$ with no evidence of either magnetic ordering or superconductivity, despite the presence of a Cu-P layer that bears a certain resemblance to the Fe-As or Fe-Se layers in the iron-based materials. Theoretical calculations suggest that, unlike other 3$d$-based magnetic materials (with a large DOS at the Fermi surface), Cu-based 122 and 265 have comparatively low Fermi-level density-of-states with the majority of the Cu spectral weight locating well below. The analyses of the structure, the DOS, the plasma frequency, and the electron-phonon coupling $\lambda$ explain the metallic behavior in the two compounds. Our results indicate the details of structures dictate characteristics of materials.

**Acknowledgement**

This work was supported by the U. S. Department of Energy (DOE), Office of Science, Basic Energy Sciences, Materials Science and Engineering Division (A.S.S.). This study was partially funded (L.L., D.P.) by ORNL's Lab-directed Research & Development (LDRD). The electron microscopy work was performed at the ORNL's Center for Nanophase Materials Sciences (CNMS), which is an Office of Science User Facility. We finally acknowledge support from the Department of Energy - National Nuclear Security Administration under Grant No. DE-NA0002014 (Y.V., G.M.T.) for pressure measurements

**Figures & Captions:**

**Figure 1.** (a) Refined powder X-ray diffraction pattern for $Ca_2Cu_6P_5$ and $CaCu_{2-\delta}P_2$. Red circles represent observed data; black and green solid lines represent the calculated intensity and difference between the observed and calculated intensity. The peak of trace impurity phase are indicated by the symbols of arrow ($CuP_2$) in $Ca_2Cu_6P_5$, "*"($CaCu_4P_2$) and "#" ($Ca_2Cu_6P_5$) in $CaCu_{2-\delta}P_2$. (b) STEM images reveal the structure of $Ca_2Cu_6P_5$ with $n = 1$ and 2 layers, and interlayer atoms, by imaging light and heavy elements simultaneously. Also, the alternating short-long Cu-Cu distances are clearly revealed.

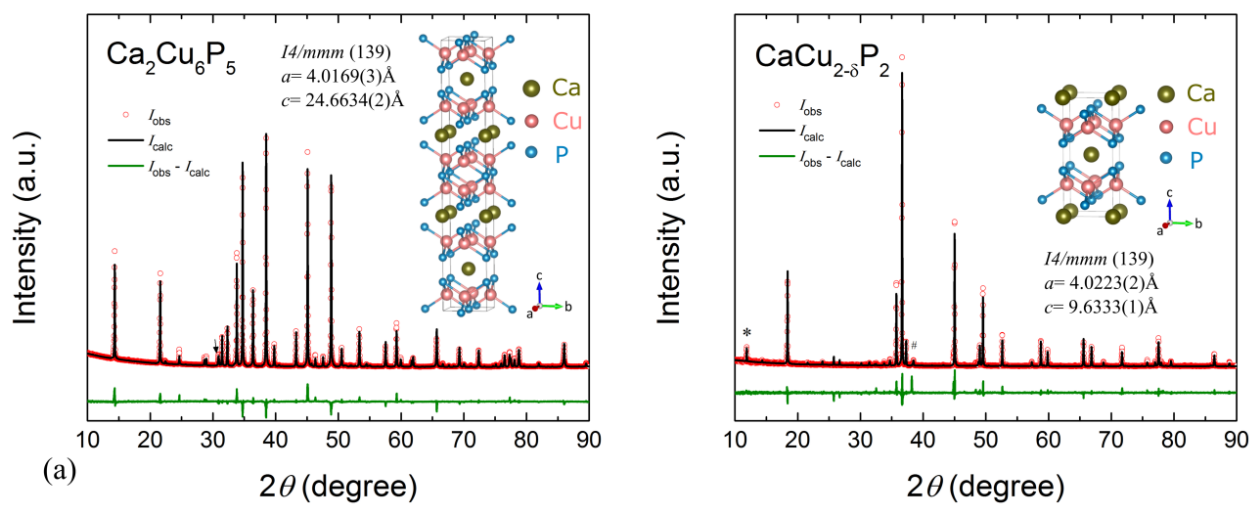

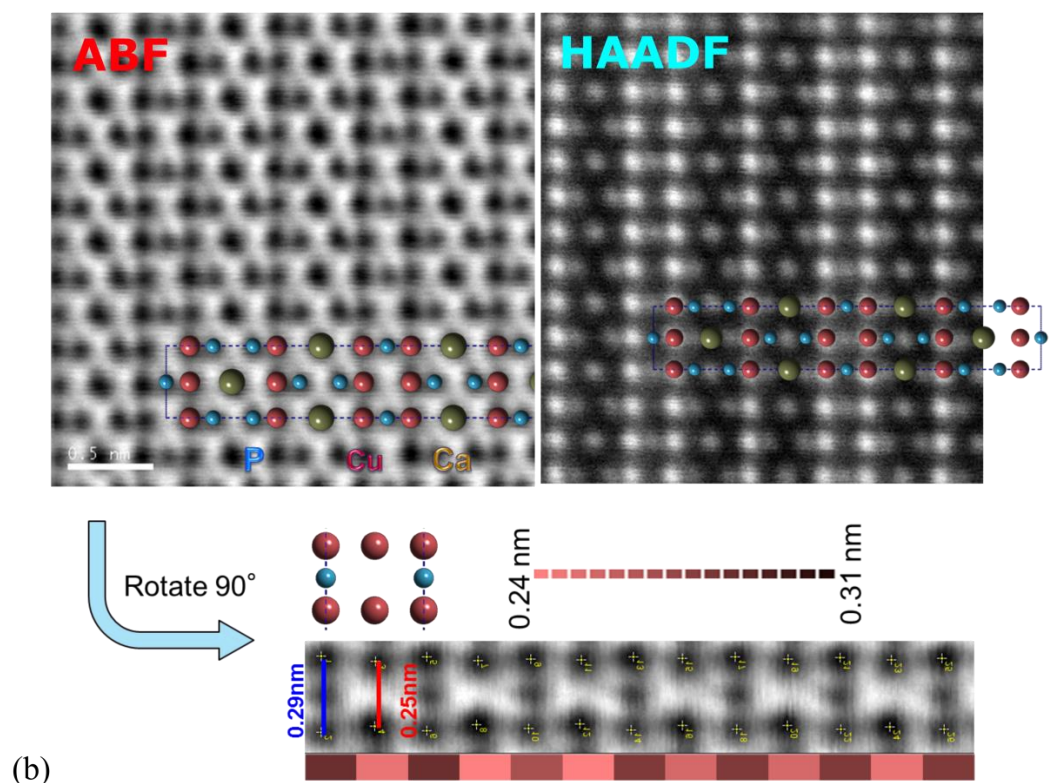

**Figure 2.** Temperature-dependence of resistivity for $Ca_2Cu_6P_5$ and $CaCu_{2-\delta}P_2$ up to room temperature. The inset is the temperature dependence of resistivity for $Ca_2Cu_6P_5$ measured below 100 K with the application of approximately hydrostatic pressure of 1.4, 3.7, 8.9, and 11.7 Gpa.

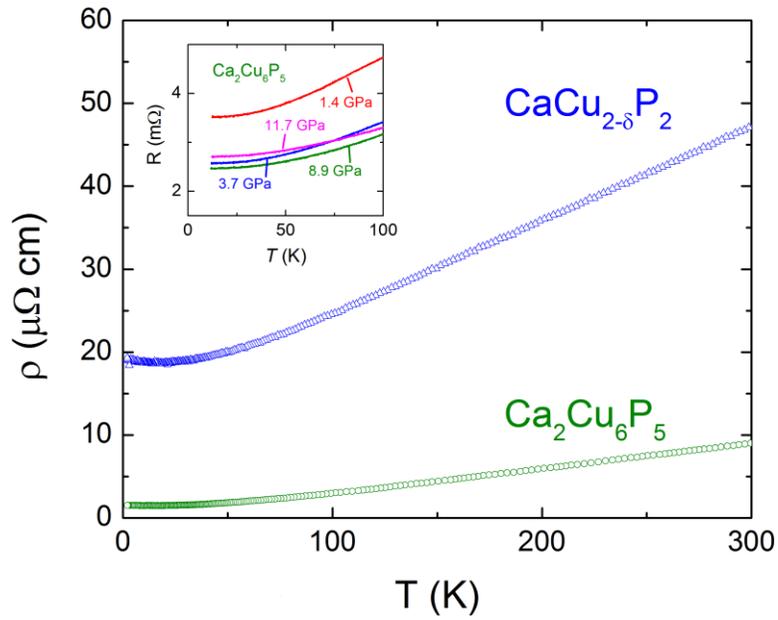

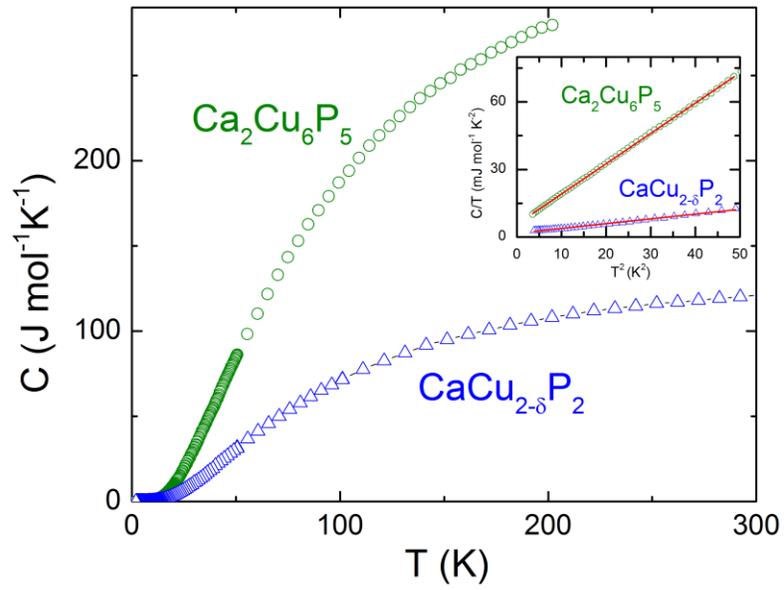

**Figure 3.** Temperature dependence of heat capacity for $Ca_2Cu_6P_5$ and $CaCu_{2-\delta}P_2$; $C/T$ versus $T^2$ in the low temperatures, along with linear fits are in inset.

**Figure 4.** Temperature dependence of the Seebeck coefficient for $Ca_2Cu_6P_5$ and $CaCu_{2-\delta}P_2$ up to room temperature; thermal conductivity is shown in the inset.

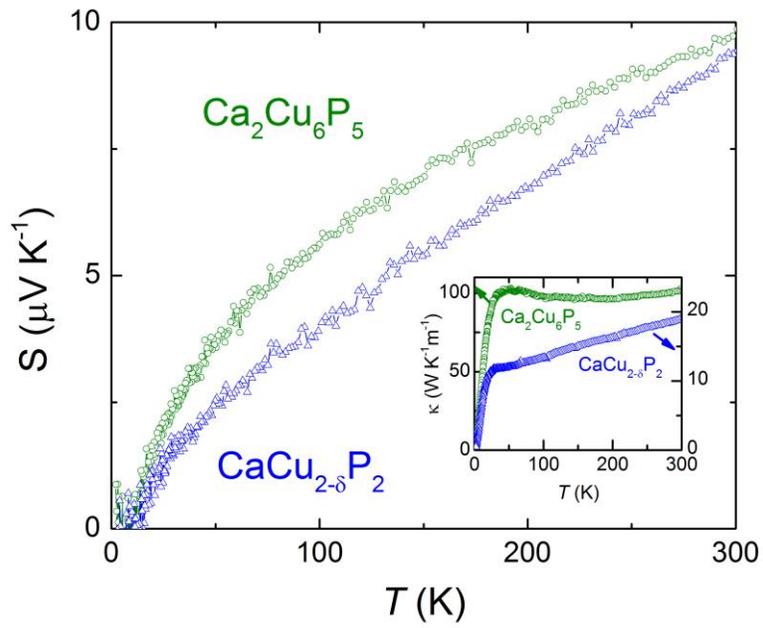

**Figure 5.** The calculated density-of-states and Fermi surface of $Ca_2Cu_6P_5$ (a) and $CaCu_2P_2$ (b). Inset of (b) shows the DOS difference between 122 (solid line) and slightly Cu-deficient cell (dotted line).

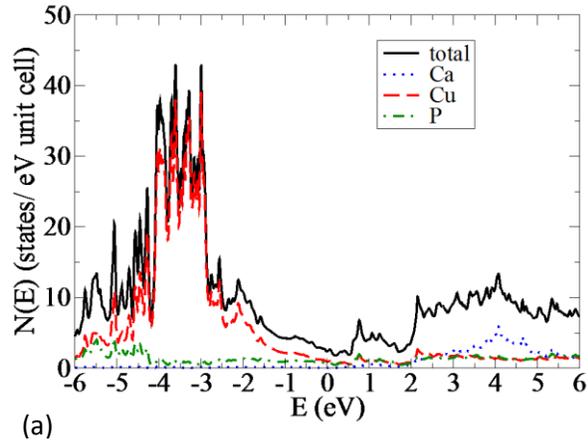
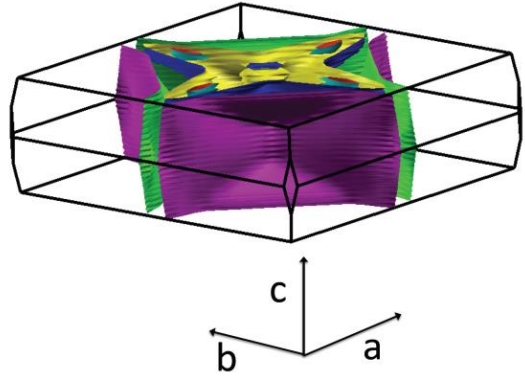

(a)

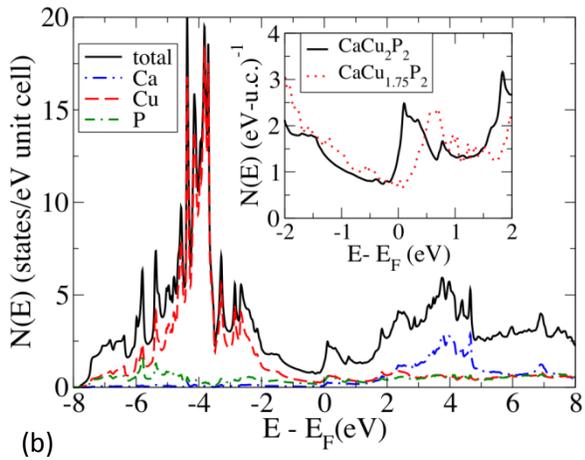
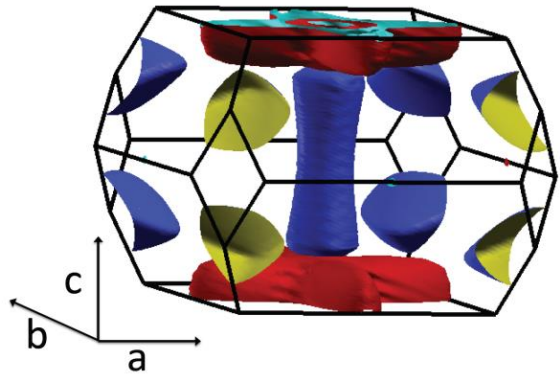

(b)